\documentclass[acmsmall]{acmart}
\usepackage{subfigure}
\usepackage{subcaption}
\usepackage{mdframed} % used in the appendix

\AtBeginDocument{%
  \providecommand\BibTeX{{%
    \normalfont B\kern-0.5em{\scshape i\kern-0.25em b}\kern-0.8em\TeX}}}

\acmYear{2024}\copyrightyear{2024}
\setcopyright{acmlicensed}
\acmConference[PECS '24]{Workshop on Performance and Energy Efficiency in Concurrent and Distributed Systems}{June 3--4, 2024}{Pisa, Italy}
\acmBooktitle{Workshop on Performance and Energy Efficiency in Concurrent and Distributed Systems (PECS '24), June 3--4, 2024, Pisa, Italy}
\acmDOI{10.1145/3659997.3660034}
\acmISBN{979-8-4007-0644-8/24/06}

\begin{document}
%%%%%%%%%%%%%%%%%%%%%%%%%%%%%%%%%%%%%%%%%%%%%
%\title[short title]{Long title}
\title[LBM energy efficiency]{Energy efficiency: a Lattice Boltzmann study}

\author{Giorgio Amati}
\orcid{0000-0003-1116-1443}
\affiliation{%
  \institution{CINECA}
  \streetaddress{Via dei Tizi 6 } % controllare!
  \city{Rome}
   \postcode{00185}  % controllare!
  \country{Italy}}
\email{g.amati@cineca.it}

\author{Matteo Turisini}
\orcid{1234-5678-9012} % controllare!
\affiliation{%
  \institution{CINECA}
  \streetaddress{Via dei Tizii 6 } % controllare!
  \city{Rome}
   \postcode{00185}  % controllare!
  \country{Italy}}
\email{m.turisini@cineca.it}

\author{Andrea Acquaviva}
\orcid{1234-5678-9012} % controllare!
\affiliation{%
  \institution{CINECA}
  \streetaddress{Via Magnanelli 2} % controllare!
  \city{Casalecchio di Reno}
   \postcode{80833}  % controllare!
  \country{Italy}}
\email{a.acquaviva@cineca.it} % controllare!

\renewcommand{\shortauthors}{Amati and Turisini, et al.}

\begin{abstract}
The energy consumption and the compute performance of a fluid dynamic code have been investigated varying parallelization approach, arithmetic precision and clock speed.
The code is based on a Lattice Boltzmann approximation, is written in Fortran and was executed on high-end GPUs of Leonardo Booster supercomputer. 
Tests were conducted on single server nodes (up to 4 GPUs in parallel).
Performance metrics like the number of operations per second and energy consumption are reported, to quantify how smart coding approach and system adjustment can contribute to reduction of energy footprint while keeping the scientific throughput almost unaltered or with acceptable level of degradation.
Results indicate that this application can be executed with $20\%$ of energy saving and reduced thermal stress, at the cost of $5\%$ more computing time.
The paper presents preliminary conclusions, as it is a first step of a larger study dedicated to energy efficiency at scale.
\end{abstract}

%
% Keywords
%





\maketitle

\section{Introduction}
Today, high-level supercomputers require tens of MW of electrical power to operate in the pre-exascale performance region and 20 MW or more are foreseen for exascale~\cite{Ref:Top500}.
Therefore, to maintain the environmental sustainability of scientific research, some form of energy optimization is critical for both code developers and data center management teams.

The growing interest in energy efficiency is demonstrated by several review articles focused on the opportunities and challenges facing high-performance computing centers, as recently documented in~\cite{nana2023energy} and~\cite{Silva2024_DecarbonizationHPC}.
Other useful reviews of models and techniques for energy saving can be found in cutting-edge research articles, e.g.~\cite{Conoci2021PowerCap} focused on maximizing the performance of multi-threaded applications in the presence of a power cap or in~\cite{Nath2015CRISP} presenting a tool for dynamic variation of the clock signal in general purpose GPUs.
On the other hand, this work has a more practical slant as it is the preliminary step of larger study dedicated to energy efficiency at scale.
It uses well-established methods of measuring the energy consumed by a single code on a single node and does so with static adjustment 
of GPU parameters.

Our attention focused on two ways of optimization: \emph{code optimization} and \emph{system tuning}.
With code optimization the original code is modified to improve the performance on a given set of hardware resources, for example changing the parallelization model for GPU deployment or changing the arithmetic precision used in computation.
System tuning, on the other hand, involves adjusting hardware settings, like processor's clock rate, to better fit the characteristics of the code.

We ran a production-grade Computational Fluid Dynamics code on single nodes of the Leonardo supercomputer, a EuroHPC JU pre-exascale machine whose architecture and components are described in~\cite{Ref:Leonardo}.
This was a first step in understanding how to simultaneously optimize a code in terms of energy and time, and to start developing a strategy for systematic optimization in large-scale runs.
The code is based on a Lattice Boltzmann Method (LBM) 
and has already been employed, using up to $O(1000)$ 
GPUs, to simulate the flow of an incompressible fluid around a complex structure called Silica Sponge
\cite{Ref:Nature}.
%\cite{Ref:Nature,Ref:EXAlb,Ref:LBM_m100}.
It has been chosen for its amenability for parallel computing and because its energy consumption was already documented in~\cite{Ref:LBMEnergy1,Ref:LBMEnergy2,Ref:Girotto}.

This paper is organized as follows: firstly the LBM technique is introduced in Section ~\ref{Sec:LBM}, then Section ~\ref{Sec:Method} describes the setup adopted for energy measurements.
In Section~\ref{Sec:Step1}, the results of different code optimizations are discussed.
Section~\ref{Sec:Step2} reports about a GPU clock scan study.
Finally, Section~\ref{Sec:Comments} provides summary, conclusions and future work.
\section{LBM in a nutshell} \label{Sec:LBM}
The Lattice Boltzmann Method (LBM) is a numerical technique to simulate fluid flow using a mesoscopic approach~\cite{Ref:succi2018}.
This method greatly simplifies the integration of Navier-Stokes equation making use of discretized space, time and velocity distributions to model the collective behaviour of particles.
In LBM, the fluid is represented by a lattice grid where each point corresponds to particles whose motion 
is determined by a finite number of velocity distributions.
The macroscopic properties of the fluid, such as velocity and density, are then inferred from these distributions evaluated on each gridpoint.
The evolution of the fluid in time is calculated in two steps: the motion of the particles within the lattice grid, called \emph{propagation}, and the 
interactions between particles in each gridpoint, called \emph{collision}.
By iterating collision and propagation steps, LBM is a simple and efficient to simulate the flow of a fluid through complex geometries and under various stress conditions. 
%LBM has gained popularity in computational fluid dynamics due to its simplicity, flexibility, and ability to parallel  computing.
Here, the LBM model is briefly sketched.

In its simplest and most compact form, called \emph{Single Relaxation Timestep} (SRT), the LBM equation reads as follows:
\begin{equation} \label{eq:eq1}
f(\vec{x}+\vec{c}_i,t+1) \equiv f'_i(\vec{x};t) = (1-\omega) f_i(\vec{x};t) + \omega f_i^{eq}(\vec{x};t) + S_i,\;\;i=1,b
\end{equation}
where $ \vec{x}$ and $\vec{c}_i$ are position vectors in ordinary space, $t$ is the time, $f_i^{eq} $ are the 
equilibrium distribution functions related to the \emph{i-th} direction, and $S_i$ is a source term.
Such an equation represents the following situation: the populations at site $\vec{x}$ at time $t$ collide 
(i.e. collision step) and produce a post-collision state $f'_i(\vec{x};t)$,
which is then scattered away to the corresponding neighbour (i.e. propagation step) at $\vec{x}+\vec{c}_i$ at time $t+1$.
The lattice time step is unitary, so $\vec{c}_i$ is the length of the link connecting a generic lattice site node 
$\vec{x}$ to its $b$ neighbors, located at $\vec{x}_i = \vec{x}+\vec{c}_i$.
For 3D simulation, a lattice with $ b=19 $ is used, hence there are $19$ directions of 
propagation (i.e. neighbours) for each grid-point $\vec{x}$.
The local equilibrium is provided by a truncation to the second order in the Mach number $M=u/c_s$, 
of the Maxwell-Boltzmann distribution, namely
\begin{equation} \label{eq:equil}
f_i^{eq}(\vec{x};t) = w_i \rho (1 + u_i + q_i)
\end{equation}
where $w_i$ is a set of weights normalized to unity, $u_i = \frac{\vec{u} \cdot \vec{c}_i}{c_s^2}$ 
and  $q_{i}=(c_{ia}c_{ib}-c_{s}^{2}\delta_{ab})u_{a}u_{b}/2c_{s}^{4}$, with $c_s$ equal to the speed of sound in the lattice, 
and an implied sum over repeated Latin indices $a,b=x,y,z$.\\
The source term $S_i$ of Eq.~\ref{eq:eq1} typically accounts for the momentum exchange between the fluid and external (or internal) fields, 
such as gravity or self-consistent forces describing the potential energy of interactions within the fluid.
Fluid dinamic quantities like density $\rho$ and velocity $ \vec{u} $ can be recovered from distributions $ f_i $:
\begin{equation} \label{eq:eq2}
\rho = \sum_i f_i  \hspace{2cm}
\vec{u}=(\sum_i f_i \vec{c}_i)/\rho,
\end{equation}
the Navier-Stokes equations for an isothermal quasi-incompressible fluid can be recovered in the continuum limit  if the lattice has the suitable symmetries and the local equilibria are chosen according to Eq.~\ref{eq:equil}.\\
Finally, the relaxation parameter $\omega$ in Eq.~\ref{eq:eq1} controls the viscosity of the lattice fluid according to
\begin{equation} \label{eq:eq3}
\nu = c_s^2 (\omega^{-1}-1/2).
\end{equation}

\begin{figure}[h]
\subfigure[OpenFoam]{\includegraphics[width=0.45\linewidth]{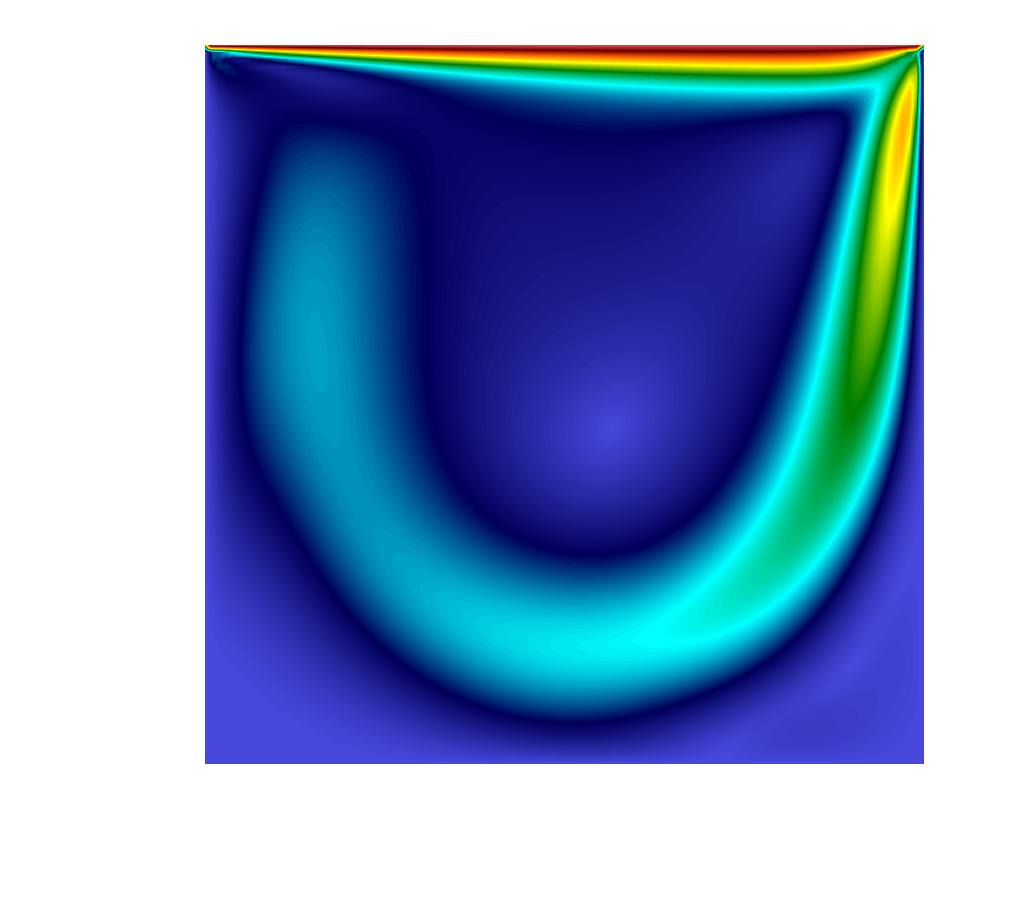}}
\subfigure[LBM]{\includegraphics[width=0.45\linewidth]{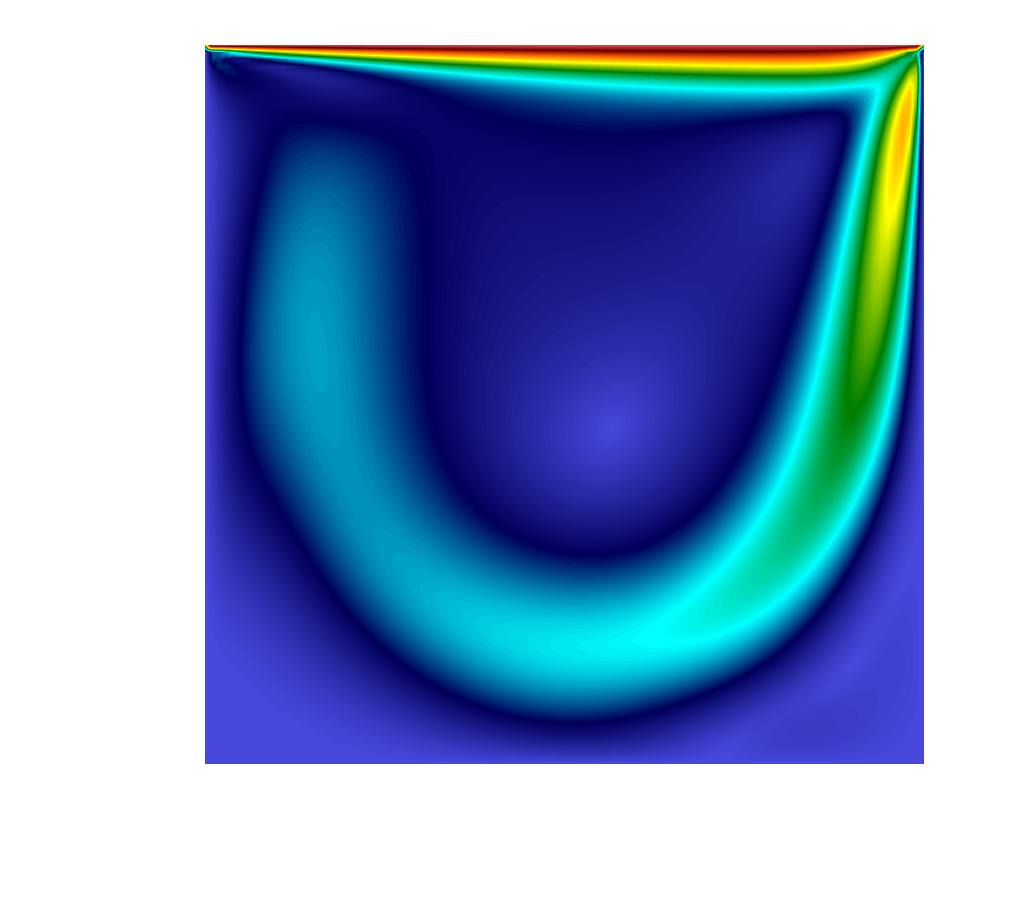}}
\caption{Lid-driven cavity velocity field distribution, qualitative comparison between OpenFoam and LBM solvers.}
\label{fig:Validation}
\Description[Lid-driven cavity velocity field distribution, qualitative comparison between OpenFoam and LBM solvers]{
Lid-driven cavity velocity field distribution, qualitative comparison between OpenFoam and LBM solvers. Both techniques, i.e. direct equation integration and lattice approximation respectively, achieve the same result. 
}
\end{figure}

\subsection{Lid-driven test case}
For this energy efficiency exploration, the selected test case is a canonical 3D lid-driven cavity, a classical benchmark problem for viscous incompressible fluid flow.
The lid-driven consist in a cubic cavity with walls with no-slip conditions and a a lid moving with a tangential velocity.
In Figure~\ref{fig:Validation} two horizontal sections of the velocity field in the fluid are presented (horizontal means perpendicular to the lid motion). 
One is obtained with the open-source OpenFoam code that directly solves Navier-Stokes equation~\cite{Ref:OF}. 
The other is calculated using our LBM code.
Both are calculated with a value $100$ for the Reynolds Number\footnote{The Reynolds number indicates the ratio between inertial and viscous forces 
in the motion generated by the different fluid's internal velocities.}.

\section{Methodology} \label{Sec:Method} 
Data were obtained with \emph{Leonardo Booster}~\cite{Ref:Leonardo}, a recent supercomputer that offers 3456 high-end servers equipped with four NVIDIA \emph{Ampere} A100 GPUs by 64~GB memory\footnote{This product can be referred to as A100-SXM4-64GB to differentiate it from commercial 
A100-SXM4-40GB and A100-SXM4-80GB. Further details can be found in ~\cite{Ref:Leonardo}.}

Using HPC jargon, we refers to GPU server as Compute Node (CN).
All tests were performed on single CNs, leaving large-scale simulation for a next phase.
Two test cases were used:
\begin{itemize}
    \item for code optimization a $16$M gridpoints lattice on single GPU  ($256 \times 256 \times 512 $)
\item for system tuning a $130$M gridpoints lattice  on the full CN ($512 \times 512 \times 512 $).
\end{itemize}
 The number of simulated time steps was 40k for both.

\subsection{Software}
The LBM code is written in \verb|Fortran90| and uses \verb|MPI| communication between processes attached to different GPUs.
From the computational point of view, the code is, like almost all CFD codes, Bandwitdh (BW) limited.
The computing performance is given in Lattice Updates Per Second (LUPS) instead of standard FLOPS (Floating point operation per second).
Both are typically expressed in units of $10^6$ (e.g. MLUPS) and are linearly related, with the conversion factor being the number of operations required for each lattice gridpoint update (250 operations in our case).

GPU metrics are accessed via registers using a vendor specific tool called \verb|nvidia-smi| (System Management Interface).
The tool is used for monitoring and for settings adjustment.
All configuration settings are static i.e. no dynamic adjustment is performed while simulation is running.
Acquired data include timestamp, instantaneous power absorption and temperature. 
Sampling rate was $ 1 $ Hz and the script to start GPU monitoring is documented in appendix~\ref{Appendix:nvsmi}.

The analysis is in C/C\texttt{++} and relies CERN's ROOT library for plotting.
Bash scripting is used for data processing as well as for automatizing the entire process, from jobs submission to logbooking and data presentation.

\subsection{Protocol} \label{Sub:protocol}
This paper presents different runs of the same LBM application.
Each run has a specific combination of coding style and GPU clock rate.
For all of them, the test consists in running the executable while recording the power consumption and other GPU metrics to characterize the GPU status (temperature, clock speed, usage).
Once the run completed, data are processed to determine the energy consumption and the elapsed time.
This two metrics are referred to as Energy-To-Solution (ETS) and Time-to-Solution (TTS), indicating they are a cost to pay to have the solution.

\begin{figure}
  \centering
  \includegraphics[width=\linewidth]{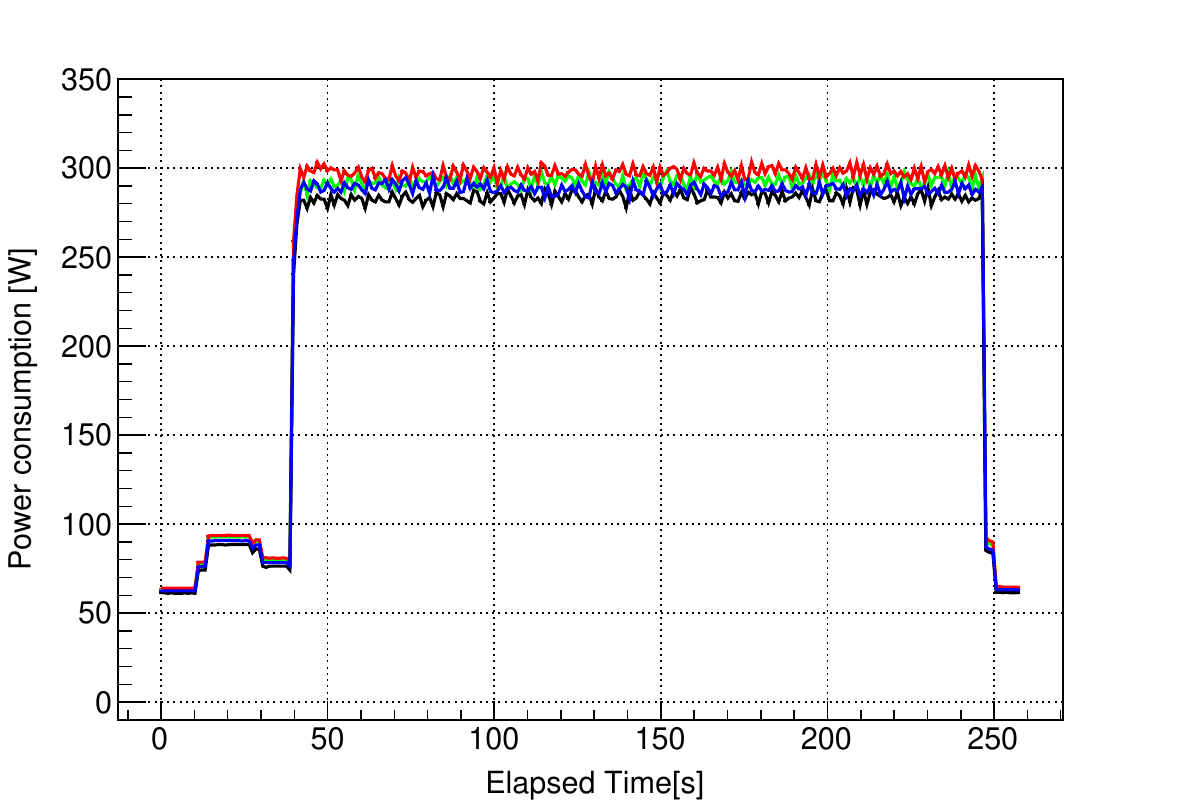}
  \caption{Power absorption profile on a single Leonardo Booster node. Colors indicate a different GPUs.}
  \label{Fig:History}
  \Description{Power profile}
\end{figure}
Figure~\ref{Fig:History} shows an example of a typical power absorption profile for a full-CN run i.e. four GPUs.
It shows a stable plateau region where the GPU is busy with the calculation, surrounded by two transient regions.
The two transient regions (before and after the plateau) are further dived in two: a transfer part in which data is moved from host to accelerator device or viceversa, and a 10-second lasting header and trailer that were introduced to characterize idle state.
Since our long term interest is on large scale simulations, all transients are neglected and the metrics are always calculated on the plateau region.
Numerically, the present example is characterized by about 200 seconds TTS and by an average power a bit lower than 300 W.
The total energy consumption (ETS), calculated as the time integral of power absorption of individual GPU power profile and then summed up at node level, is 240~kJ in this example.
\section{Code optimization} \label{Sec:Step1}
The LBM code documented in \cite{Ref:EXAlb,Ref:LBM_m100} is our starting point and uses double precision arithmetic (FP64, baseline version).
Here, a recap of its main features is presented (see Section~\ref{Sec:LBM} for model's variables).
\begin{itemize}
    \item \emph{Overhead reduction}: The code uses no functions. Explicit inlining is performed instead.
    \item \emph{Spatial locality}: local arrays are used to minimize the number of memory accesses i.e. to avoid redundant I/O operations.
    \item \emph{Memory access optimization}: only population $f_i$ are stored in memory. All other {\it derived} quantities like $\rho $ and $ \vec{u}$ are computed {\it on the fly}.
    \item \emph{Loops merging}: whenever possible, the loops were merged to reduce the number of independent kernels to be offloaded to the GPUs.  
    \item \emph{Common subexpression elimination}: eliminate redundant calculations. As an example to compute $\rho $ and $ \vec{u}$ from eq.\ref{eq:eq3} there are no multiplications to perform, because direction $ \vec{c}_i $ can assume only $ -1,0,+1$ values.
\end{itemize}
%%%%%%%%%%%%%%%%%%%%%%%%%%%%%%%%%%%%%%%%%%%%%%%%%%%%%%%%%
\subsection{Parallelization directives} \label{Sub:directives}
%%%%%%%%%%%%%%%%%%%%%%%%%%%%%%%%%%%%%%%%%%%%%%%%%%%%%%%%%
When dealing with GPU acceleration, the initial step usually is the choice of a parallelization method.
We opted for the simple and portable approach offered by directive-based parallelization models, instead of a low-level GPU coding. 
In \verb|Fortran| language there are three possible models: 
\begin{itemize}
    \item \verb|do concurrent| is a statement introduced by Fortran 2008 standard~\cite{Ref:DC}. 
    Under certain conditions, it causes loop complete unrolling that means that loop-iterations are executed as concurrently as possible on available hardware resources. 
    \item \verb|OpenACC| is a directive-based programming model designed for GPU offloading~\cite{Ref:openacc}. 
    It consists in decorating the code with preprocessor directives to tell the compiler how to handle parallelizable portions of the code. 
    Its usage is straightforward and does not require knowledge of low level programming techniques.
    \item \verb|OpenMP offload|: is an extension to GPU devices of the OpenMP parallelization model that was originally designed for 
    multi-core CPUs~\cite{Ref:OpenMP}. The offload to GPU is completely transparent to the programmer, with high-level syntax directives similar to OpenACC.
\end{itemize}
%
%%%%%%%%%%%%%%%%%%%%%%%%%%%%%%%%%%%%%%%%%%%
\begin{table}
 \caption{Results for different parallelization (single GPU)}
   \label{Tab:ParallelModel}
  \begin{tabular}{cccc} \toprule
       Model               & Energy & Time   &Performance\\ 
                           &  [kJ]  & [s]    & [MLUPS]   \\ \midrule
    \verb|DoConcurrent|    & 200    &  735   &  1903  \\
    \verb|OpenACC|        & 195    &  736   &  1900  \\ 
    \verb|OpenMP offload|  & 208    &  730   &  1925  \\  \bottomrule
\end{tabular}
\end{table}
%%%%%%%%%%%%%%%%%%%%%%%%%%%%%%%%%%%%%%%%%
These three models were tested on single GPUs using \verb|nvfortran| compiler rel. 23.1 and results are shown in Table~\ref{Tab:ParallelModel}.
The test was repeated on $ 20 $ different nodes exhibiting the same performance within a range 
below $ 0.2\% $.
The performance appears to be model independent, indicating substantial equivalence between the three.
This is in accordance with what is documented in~\cite{Ref:DoConcurrent}.
In the rest of the work, the \verb|OpenACC| model was chosen because it has the highest energy saving at the cost of a tiny performance loss compared to the others.

\subsection{Fused implementation}
The \emph{roofline} model \cite{Ref:Roof} is typically used to determine how close a simulation is to the maximum achievable performance on a machine, considering its nominal memory bandwidth and computing performance. 
In a code, the link between these two characteristics is its \emph{arithmetic intensity} (AI) defined as the ratio between the number of operation to be performed (additions and multiplications) and the amount of information to be moved (read and write).
Our baseline code requires $250$ operations and $ 592 $ bytes to be transferred for each grid point in the lattice, thus its arithmetic intensity is $0.42$.
By inverting the definition of AI and given a nominal bandwidth of $ 1600 $ GB/s on Leonardo's GPUs, the maximum achievable performance of our code is $670 $ GFLOPS, corresponding to $2700 $ MLUPs, using LBM jargon.
It is worth to notice that the single GPU test, reported in Table~\ref{Tab:ParallelModel}, reaches about $70\%$ of the nominal compute power, indicating that the baseline code is already quite efficient at this arithmetic intensity.

To improve the compute efficiency i.e. increase AI, we fuse the two routines of the baseline implementation 
(propagation and collision, see Section\ref{Sec:LBM}), since the first copies data from grid points to neighbours (data movement only) while the second is completely local and compute intensive.
This coding approach is referred to \emph{fused implementation} and is based on pointers swapping instead of memory I/O.
This double the memory footprint (largely affordable), halved the bandwidth requirement and, most important, double the arithmetic intensity up to $0.84$.
As a consequence, the fused implementation turns the nominal performance limit on single GPU to $5400$ MLUPS and on the full CN to $21600$ MLUPS.

%%%%%%%%%%%%%%%%%%%%%%%%%%%%%%%%%%%%%%%%%%%%%%5
\begin{table}[h]
  \caption{Results for fused implementation (Full-CN)}
   \label{Tab:Implementation}
  \begin{tabular}{cccc} \toprule
       Model               & Energy & Time &Performance\\ 
                          &  [kJ]  & [s]      & [MLUPS]   \\ \midrule
    Baseline   & 847 & 797     &  7281  \\
    Fused      & 436 & 432     & 14413  \\  \bottomrule
\end{tabular}
\end{table}

%%%%%%%%%%%%%%%%%%%%%%%%%%%%%%%%%%%%%%%%%%%%%%%%%%%%%%%%
Test results are in accordance with the factor $2x$ estimation and are presented in  Table~\ref{Tab:Implementation}.
In the remainder of the article the fused implementation parallelized with \verb|OpenACC| is used.

\subsection{Arithmetic precision}
One of the most promising optimization technique involves the arithmetic precision adopted in computation and for storage.
As a matter of facts, as the data size shrinks, operations become faster and consume less energy.
So, provided that arithmetic limits do not introduce a significant deterioration in simulation results (this must be tailored on application basis), the reduced precision is a very promising candidate to save both energy and time.
Different combinations of double (FP64) and single (FP32) precision have been explored:
\begin{itemize}
    \item \emph{double}: distribution functions are stored using arrays in double precision and 
    all computation is performed in double precision using scalar variables. 
    This precision presents an AI of $0.84$ and is the one used in the baseline version of the code.
    \item \emph{single}: distribution functions are stored using arrays in single precision and computation is performed in single precision using scalar variables. 
    This precision presents an AI of $1.68 $, rising the theoretical performance limit for the whole CN up to 
    $43000$ MLUPS. 
    \item \emph{mixed}: distribution functions are stored using arrays in single precision and all computation is performed in double precision using scalar variables.
    This precision presents the same AI as the \emph{single} case. 
\end{itemize} 

%%%%%%%%%%%%%%%%%%%%%%%%%%%%%%%%%%%%%%%%%%%%%%%%%5
%
\begin{table}[ht]
  \caption{Results for different arithmetic (Full-CN)}
  \label{Tab:Precision}
  \begin{tabular}{cccc} \toprule
    Precision & Energy  & Time    &Performance\\ 
              &  [kJ]   & [s]     & [MLUPS]   \\ \midrule
    double    & 436     & 432     & 14413     \\
    mixed     & 290     & 269     & 25330     \\   
    single    & 252     & 258     & 25756     \\
 \bottomrule
\end{tabular}
\end{table}
%
%%%%%%%%%%%%%%%%%%%%%%%%%%%%%%%%%%%%%%%%%%%%%%%%%%5
Test results are presented in Table ~\ref{Tab:Precision}.
As expected from arithmetic intensity consideration, the \emph{single} or \emph{mixed} precision show a significant improvement with respect to \emph{double}. 
In absolute terms, the \emph{single} precision has the best energy and time values, but in order to preserve generality (reduced rounding error) the \emph{mixed} precision seems to be an equivalent candidate, therefore, for the reminder of the work the mixed precision was used on top of the fused implementation parallelized with \verb|OpenACC|.
%According to the \emph{roofline} model this choice allows for $ \simeq 60\% $ of the maximum theoretical performance.

\section{System tuning}  \label{Sec:Step2}
In any processor, the clock rate of arithmetic units impacts linearly on the performance of compute-bounded codes, so the higher the clock speed, the faster execution is obtained.
Unfortunately the power consumption goes quadratically with the clock frequency, so, the energy absorption and the thermal stress on the system go up dramatically as the clock rate rises.
On the other hands, for bandwidth-limited codes, the relation is unclear since the communication process is the performance bottleneck and the power consumption of arithmetic units is minimum when those units are not fed with data.
Therefore, to clarify this aspect and quantify the energy performance of our LBM code, a clock rate scan was performed.

Before discussing the results, two element must be added to the setup description in Section~\ref{Sec:Method}, for completeness.
Firstly, the memory clock rate of our GPU model is fixed (to $1593$ MHz), so variations cannot be explored. As a consequence, only the clock rate of the streaming multiprocessor\footnote{Streaming multiprocessor are the GPU arithmetic logic units in the vendor's idiom} (SM) is available for adjustment.
Secondly, the \emph{default} configuration on our platform has an automatic management of the SM clock speed: if some thermal and power conditions are met, the base clock ($1245$ MHz) is risen to the maximum ($1395$ MHz).
Now, since on Leonardo the direct liquid system cooling is very efficient, those conditions are always met, so the default condition for us $1395$ MHz.
For this study however, it was verified that when the user (with special permissions) set a specific clock rate for the SM, the \emph{autoboost} feature is disabled and the clock is steady at the specified setting.

%This guarantees the best TTS for compute-bounded application, but can results sub-optimal for bandwidth limited code, especially considering the associated energy consumption, as we are going to see shortly.

\begin{figure}[ht]
 \centering
 \includegraphics[width=\linewidth]{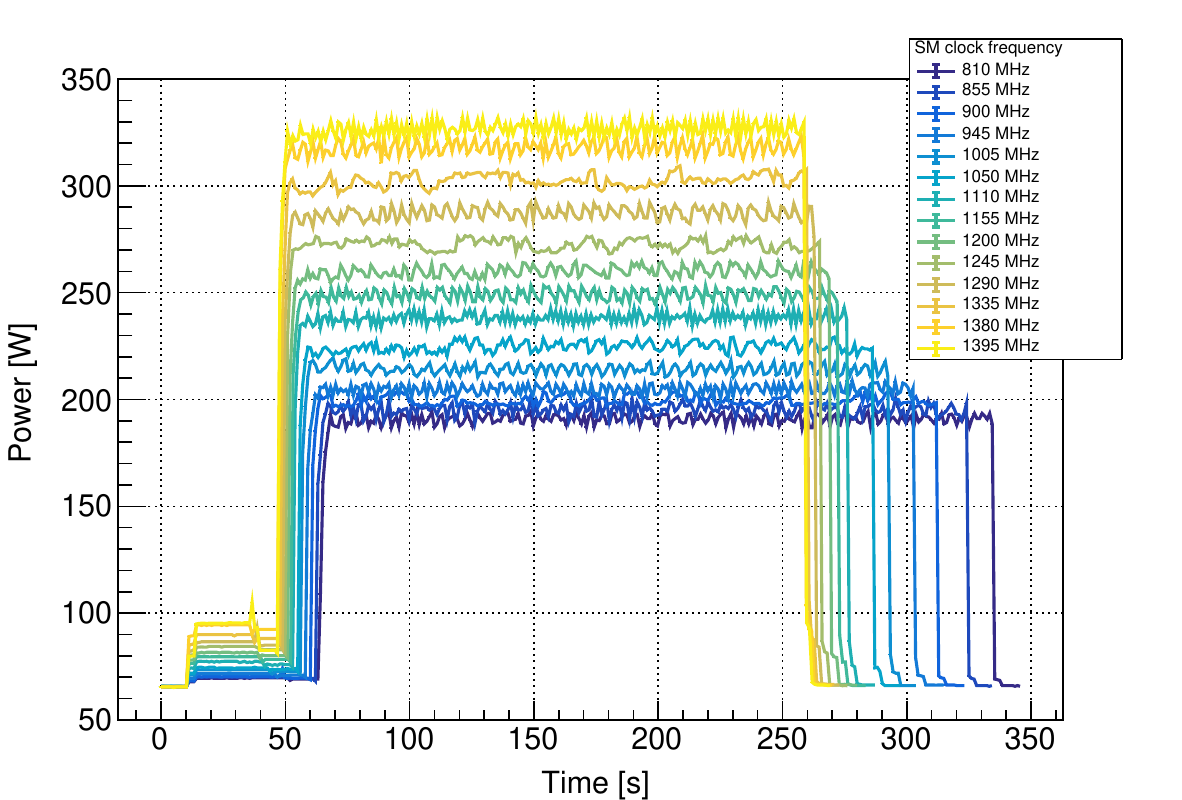} 
 \caption{Power profile for different SM clock rates}
 \label{Fig:SMFreq}
 \Description[Single-GPU power profile for different SM clock rates]{The figure shows the power absorption as a function of time for a single GPU running the LBM code at different clock speed setting.}
\end{figure}
%
%%%%%%%%%%%%%%%%%%%%%%%%%%
\begin{table}%[h]
  \caption{Results for different clock rate (Full-CN)}
  \label{Tab:FreqSM}
  \begin{tabular}{cccc} \toprule
    Clock frequency & Energy  & Time    & Performance \\
    MHz & [kJ]  & [s]    & [MLUPS] \\ \midrule   
810 & 223 & 346 & 19829\\
855 & 220 & 334 & 20461\\
900 & 217 & 323 & 21265\\
945 & 216 & 315 & 21894\\
1005& 218 & 304 & 22640\\
1050& 223 & 297 & 23160\\
1110& 228 & 287 & 24011\\
1155& 235 & 284 & 24317\\
1200& 242 & 280 & 24565\\
1245& 250 & 276 & 24886\\
1290& 260 & 273 & 25125\\
1335& 272 & 271 & 25277\\
1380& 286 & 270 & 25313\\
1395& 290 & 269 & 25337\\  \bottomrule
\end{tabular}
\end{table}
% ./print_result.sh bgk3432.txt | awk '{print $4,$5,$6,$8}'
%%%%%%%%%%%%%%%%%%%%%%%%%
For this test, three nodes have been randomly selected for reproducibility of the results and to derive some indications on variability inside the cluster for future large scale test.
As shown in Table~\ref{Tab:FreqSM} data were collected for 14 different clock rates spanning from 810 MHz to $1380$ MHz plus our default condition that correspond to $1395$ MHz as stated above.
The corresponding power profiles are shown with different color scale in Figure~\ref{Fig:SMFreq} for one of the GPUs running LMB on the full-CN.

As expected, the relationship between clock rate and TTS is monotonic on the entire range: the faster the first, the shorter the second.
This does not hold for energy, which shows a U-shape dependency on clock.
This is a strong indication that, for this application, the maximum clock speed is not the highest energy efficiency work point.

\begin{figure}[ht]
 \centering
  \includegraphics[width=\linewidth]{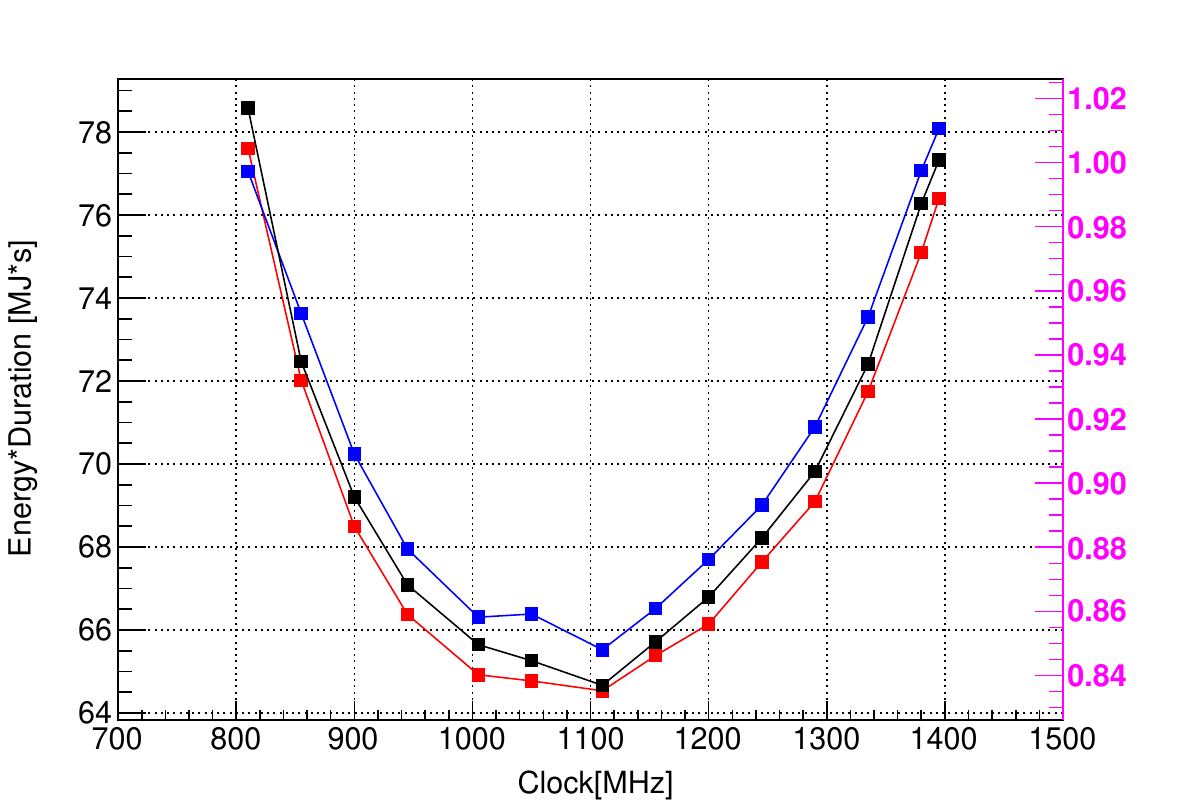}
 \caption{Product \emph{energy $\times$ duration} as function of GPU clock speed for three different compute nodes. Right y-axis indicates normalized values.}
 \label{Fig:GAIN}
  \Description[The product \emph{energy $\times$ duration} as function of the clock speed for three compute nodes. Right y-axis indicates normalized values.]{The product \emph{energy $\times$ duration} as function of the clock speed for three compute nodes. Right y-axis indicates normalized values. The curve is U-shaped with a minimum between 1000 MHz and 1100 MHz. The maximum value of the curves is at the level 78 MJoules $\times$ seconds while the minimum value is $15\%$ less.}

\end{figure}

To better quantify this statement (monotonic time and u-shaped energy) we introduced two quantities.
The first, shown in Figure~\ref{Fig:GAIN} as a function of the clock rate, is their product: energy $\times$ time in units of \emph{Joules $\times$ seconds} (called an \emph{action} in Physics).
Since we want to minimize simultaneously the energy consumption and the elapsed time, the minimum value of their product is the best work point.
Clearly, the \emph{action} curve in Figure~\ref{Fig:GAIN} is U-shaped, indicating an optimal clock region between 1000 MHz and 1100 MHz, for all the three nodes considered.
Moreover, variations between CN (different color used for different CN) can be used as a measure of the total error resulting from CN variability and measurement sensitivity.
Using the average value in units of Joule$\times$second$\times 10^6$ for the maximum clock rate as normalization, the total error is at the level of $2\%$ percent.

\begin{figure}[h]
 \centering
 \includegraphics[width=\linewidth]{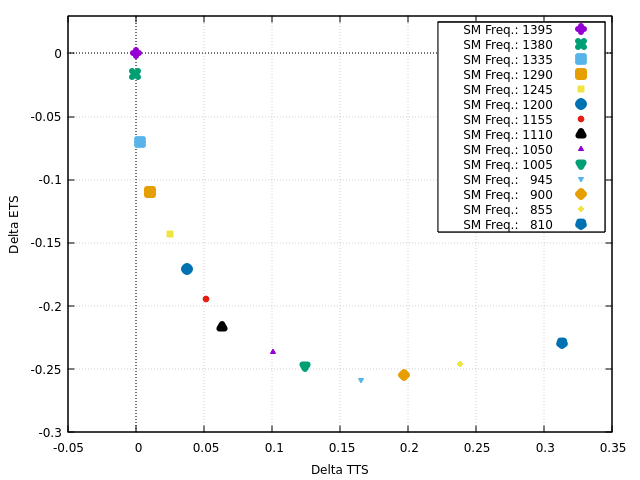}
 \caption{Normalized variations of ETS and TTS with respect to those obtained at the maximum clock rate.}
 \label{Fig:ETS}

\Description[Normalized variations of ETS and TTS with respect to those obtained at the maximum clock rate.]{The figure is a scatter plot of Energy vs Time. Each clock speed setting is a point. The coordinates of each point are the normalized variations with respect to the maximum clock rate.
Thus the point corresponding to the maximum clock rate is at the origin of this reference system.}

\end{figure}

For clearer view, a second representation was introduced, plotting the normalized variation of ETS and TTS with respect to the maximum clock rate on the 2D plane of Figure~\ref{Fig:ETS}.
In this representation, the origin is the default configuration and the different clock spread on the plane indicating a normalized combination of energy and time.
In particular, since our default is the maximum clock speed, the energy variation takes negative values only (on the y-axis).
Moving along the y-axis ($x\simeq0$), it is worth to notice that a  negligible time performance degradation can be recognized down to $1290$ MHz where the TTS variation is at the level of $1\%$ and the energy saving is $10\%$.

\begin{figure}[hb]
 \centering
\includegraphics[width=\linewidth]{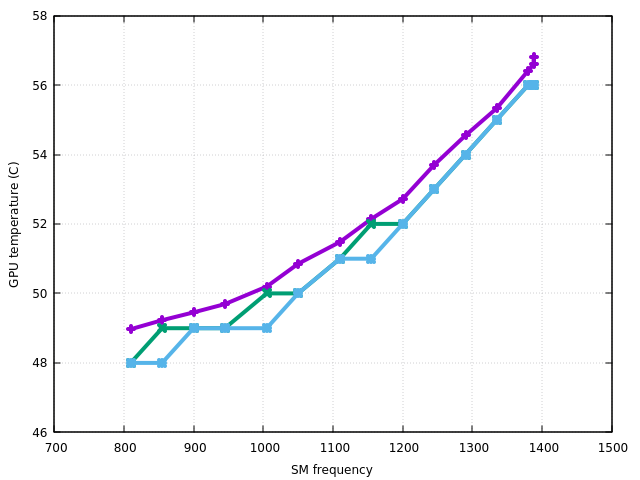}
 \caption{GPU average temperature as a function of the SM clock speed for three compute nodes.}
 \label{Fig:Temperature}
 \Description[GPU average temperature as a function of the SM clock speed for three compute nodes.]{GPU average temperature as a function of the SM clock speed for three compute nodes. This demonstrates a monotonic relationship between clock rate and temperature of the GPU: the higher the first the higher the second.}
\end{figure}

Moreover, an interesting effect of clock tuning is that, with energy, the average power consumption also decreases and so does the thermal stress on GPUs. 
In fact, a significant variation of the average temperature by few Celsius degrees was recorded during the run and is presented on Figure~\ref{Fig:Temperature} as a function of the clock speed for the three CNs.
In our view, this is a positive effect since a decreased temperature can reduce the number of hardware failures and increase system's resilience and up-time, especially for large-scale infrastructures.
\section{Summary, Comments and Future work} \label{Sec:Comments}
The energy saving of a fluid dynamic code was investigated on few single nodes of the GPU partition of Leonardo supercomputer.
The study, performed on a lattice approximation,  indicates that an optimal work point for GPU clock does exist ($1155$ MHz) for this particular code and it allows to save about $20\%$ of energy at the price of $5\%$ more elapsed time.
In general, test results can be used to calibrate the running mode between maximum energy saving and fastest execution.

In this last part of the paper, some comments are presented with the help of Table~\ref{Tab:Recap} that recap some relevant steps in our tuning and express the \emph{energy cost} of a computation in terms of $Joule/MLUPS$.

%%%%%%%%%%%%%%%%%%%%%%%%%%%%%%%%%%%%%%%%55
\begin{table}
     \caption{Relevant points in optimization flow}
    \label{Tab:Recap}
  \begin{tabular}{cccc} \toprule
    Step            & Energy  & Time  & Operation cost \\ 
                    &  [kJ]   & [s]   & [J/MLUP]   \\ \midrule
    baseline        &  821    &  797  &  153 \\ 
    fused           &  436    &  432  &  81 \\ 
    mixed precision &  290    &  269  &  54 \\ 
    SM clock=1290   &  260    &  273  &  48 \\ 
    SM clock=1155   &  235    &  284  &  45 \\ 
    SM clock=1005   &  218    &  304  &  41 \\     \bottomrule
\end{tabular}
\end{table}
%%%%%%%%%%%%%%%%%%%%%%%%%%%%5555 

\subsection{Code Optimization} \label{recap_code_optimization}
Source code adjustment determined in our case both a reduction in TTS and an improvement in ETS as well.
A summary of our observations is presented here:
\begin{itemize}
    \item \emph{GPU offloading}: the three parallelization models under test exhibit similar performance. 
    \verb|OpenMP offload| is slightly faster than \verb|openACC|, but requires $6\%$ percent more energy.
    \item \emph{Performance limit}: based on the GPU specifications, since LBM is a bandwidth-limited code, a $ \simeq 70\% $ of the theoretical peak performance has been achieved just with code optimization.
    \item \emph{Implementation}: With the fused implementation (i.e. code rethinking), ETS and TTS are reduced by almost a factor $2x$.
    \item \emph{Arithmetic precision}: single precision exhibits an almost linear speed up and energy saving with respect to double precision. The gain was $ 1.67x $ for TTS and $ 1.73x $ for ETS.
    The \emph{mixed} precision have similar slightly worse performance, but it is a viable option whenever the double precision is not mandatory. A further factor $2x$ could be achieved using mixed half-single precision, 
    but this feature was not explored, cause is too test-case dependent.  
\end{itemize}
At the end of the code optimization step, the best combination is the one with \emph{fused} implementation, \emph{mixed} precision and using \verb|openACC| model for parallelization on GPUs.
This combination reduces energy and time by the same factor, with respect to the baseline version: i.e.  $ \simeq 2.83x$ energy saving and $\simeq 2.96$ latency reduction.
This reduction in energy consumption takes into account also the overhead caused by MPI, that is $ \simeq 10 \% $ of the total time.

\subsection{System Tuning}
The impact of GPU clock rate on application's TTS and ETS has been investigated using the optimal combination described in \ref{recap_code_optimization}.
The investigation was static, meaning that no dynamic control mechanism was active during the execution of the test.
As expected, clock speed reduction determines TTS increase.
However, in the range from $1395$ MHz down to $ 1005 $ MHz,the energy decreases more rapidly than time, indicating a benefit for energy consumption.
As an example, at $1290$ MHz an energy saving of $10\%$ has been measured together with a negligible $1\%$ increase in elapsed time.
Further reduction of the clock speed (below 1000 MHz) causes no further energy saving. On the contrary the energy cost of the computation increases.
%%%%%%%%%%%%%%%%%%%%%%%%%%%%%%%%%%%%%%%%%%%%%%%%
\subsection{Final Remarks and Future work}
%%%%%%%%%%%%%%%%%%%%%%%%%%%%%%%%%%%%%%%%%%%
Three major findings that can be pinpoint from this preliminary work are:
\begin{itemize}
    \item Significant energy and time saving is associated with code optimization.
    At this level, energy consumption and time seems linearly related.
    So, the code optimizations should be seen as the first and most crucial step for energy efficiency.
    In our case, the measured energy cost reduction is $3x$ as reported in Table~\ref{Tab:Recap} from $153$ to $54$ J/MLUPS.
    
    \item When tuning the system, the choice between saving energy or time or something in the middle, heavily depends on many factors, like the kind of application or the specific test-case.
    However, it is reasonable to suppose that for the same class of codes (e.g. LBM codes) or a general group, 
    like CFD BW-limited codes, the ratio costs versus benefits would be the similar, once the code optimization step is fulfilled. 
    \item Using lower clock rate on GPU,  the device operates at lower temperature, lessening thermal stress on the system, since it is well known that the number of hardware failures increases as the temperature increases.
\end{itemize}

Another interesting consideration is that by projecting a few percent of energy saving on a facility consuming MW of electrical power, the saving in energy cost would be serious and so the CO$_2$ emission~\cite{Ref:CO2}.
For example, running our LBM at $ 1200 $MHz would reduce the power requirements by $ \simeq 50 $ W for GPU at the acceptable cost of  increased run duration by $ 4\% $. 
This means that a run using the whole Leonardo Booster partition would ideally imply a power saving of $\simeq 0.8$MW, which would be a huge and good saving.
In practice, parallel applications running at scale, on multiple nodes, require a more refined analysis and system tuning to obtain a full energy scalability, so the actual saving should be evaluated case by case.

Motivated by these considerations, we plan to extend the investigation to multi-node simulations and in particular perform the following studies:
\begin{itemize}
    \item Test other applications on Leonardo and possibly test dynamic adjustment tools.
    \item Scale up LBM simulation to $ O(100/1000)$ GPUs.
    \item Test LBM using other platforms and GPU models.
\end{itemize}

% ACKNOWLEDGEMENT
\begin{acks}
Thanks to Federico Tesser and Giuseppe Palumbo from CINECA for their help in the initial phase of this work.
\end{acks}

% BIBLIOGRAFY
\bibliographystyle{ACM-Reference-Format}
\bibliography{biblio}

% APPENDIX
\appendix
\section{nvidia-smi script}\label{Appendix:nvsmi}
To start the monitoring of all GPUs on a compute server node, the following script was used.
To stop measurements the \verb|pkill| command was used.\\
%\vspace{0.5cm}

\begin{mdframed} {\small
\begin{verbatim}
#!/bin/bash 
# set up monitoring and start measurements 
node=$(hostname | awk 'BEGIN { FS = "." } ; {print $1}') 

queries=""  # define queries
queries+="index" 
queries+=",timestamp" 
queries+=",power.draw" 
queries+=",clocks.sm" 
queries+=",clocks.mem" 
queries+=",temperature.gpu" 
queries+=",temperature.memory" 
queries+=",utilization.gpu" 
queries+=",utilization.memory" 

format=""  # Define format 
format+="csv" 
format+=",noheader" 
format+=",nounits" 

period=1000  # # Define period in millisecond
# Start sampling 
nvidia-smi --query-gpu=$queries \ 
           --format=$format -lms $period \ 
           >> $node.nvidiasmi.txt 2>&1 & 
#eof
\end{verbatim}} \end{mdframed}

\end{document}